\documentclass[trackchanges,resetfootnote]{aastex701}
\usepackage{amsmath}
\usepackage{capt-of}   
\usepackage{empheq}
\usepackage{enumitem}
\usepackage{hyperref}
\usepackage{hhline}
\usepackage{numprint}
\usepackage{pifont}
\usepackage{xspace}
\usepackage{placeins}
\usepackage{graphicx}
\usepackage{mwe}
\usepackage{multirow}
\usepackage{tabularx}
\usepackage{changepage}
\usepackage{float}
\usepackage{booktabs} 
\usepackage{array} 
\usepackage{amsmath}
\usepackage{amssymb}
\usepackage{threeparttable}

\defcitealias{ecker2026}{E26}

\newcommand{\threecodes}{\texttt{lenstronomy} and \textsc{Glee}}

\begin{document}


\title{Supernova 2025wny: High-angular resolution Keck/NIRC2 observations and preliminary lens modeling}

\author[orcid=0000-0002-0385-0014]{Christopher J. Storfer}
\affiliation{Institute for Astronomy, University of Hawai'i at Manoa, 2680 Woodlawn Dr., Hawai'i, HI 96822, USA }
\email[show]{cstorfer@hawaii.edu}  

\author[orcid=0000-0002-8459-7793]{Kenneth C. Wong}
\affiliation{Research Center for the Early Universe, Graduate School of Science, The University of Tokyo, 7-3-1 Hongo, Bunkyo-ku, Tokyo 113-0033, Japan}
\email{kcwong19@gmail.com}  

\author[orcid=0000-0003-3108-9039]{Ana Acebron}
\affiliation{Instituto de Física de Cantabria (CSIC-UC), Avda. Los Castros s/n, 39005 Santander, Spain}
\affiliation{INAF -- IASF Milano, via A. Corti 12, I-20133 Milano, Italy}
\email{ana.acebron@unican.es}  

\author[orcid=0000-0002-5926-7143]{Claudio Grillo}
\affiliation{Dipartimento di Fisica, Universit\`a  degli Studi di Milano, via Celoria 16, I-20133 Milano, Italy}
\affiliation{INAF -- IASF Milano, via A. Corti 12, I-20133 Milano, Italy}
\email{claudio.grillo@unimi.it} 

\author[0000-0003-3953-9532]{Willem~B.~Hoogendam}
\altaffiliation{NSF Graduate Research Fellow}
\affiliation{Institute for Astronomy, University of Hawai'i at Manoa, 2680 Woodlawn Dr., Hawai'i, HI 96822, USA }
\email{willemh@hawaii.edu} 

\author[orcid=0000-0001-8156-0330]{Xiaosheng Huang}
\affiliation{Department of Physics \& Astronomy, University of San Francisco, San Francisco, CA 94117-1080}
\affiliation{Physics Division, Lawrence Berkeley National Laboratory, 1 Cyclotron Road, Berkeley, CA, 94720}
\email{}

\author[0000-0002-6230-0151]{David~O.~Jones}
\affiliation{Institute for Astronomy, University of Hawai'i, 640 N. A'ohoku Pl., Hilo, HI 96720, USA}
\email{dojones@hawaii.edu}

\author[orcid=0000-0002-7965-2815]{Eugene A. Magnier}
\affiliation{Institute for Astronomy, University of Hawai'i at Manoa, 2680 Woodlawn Dr., Hawai'i, HI 96822, USA }
\email{}

\author[0000-0001-9846-4417]{Kaisey~S.~Mandel}
\affiliation{Institute of Astronomy and Kavli Institute for Cosmology, University of Cambridge, Madingley Road, Cambridge, CB3 0HA, UK}
\email{}

\author[orcid=0009-0009-8206-0325]{Nicolas Ratier-Werbin}
\affiliation{Physics Division, Lawrence Berkeley National Laboratory, 1 Cyclotron Road, Berkeley, CA, 94720}
\affiliation{Department of Physics, Complutense University of Madrid, 28040 Madrid, Spain}
\email{}

\author[orcid=0000-0001-5402-4647]{David Rubin}
\affiliation{Department of Physics and Astronomy, University of Hawai`i at Mānoa, Honolulu, Hawai`i 96822}
\email{}

\author[0000-0003-4631-1149]{Benjamin~J.~Shappee}
\affiliation{Institute for Astronomy, University of Hawai'i at Manoa, 2680 Woodlawn Dr., Hawai'i, HI 96822, USA }
\email{shappee@hawaii.edu}

\author[orcid=0009-0004-0717-6268]{Oscar Soler-Perez}
\affiliation{Department of Physics \& Astronomy, University of San Francisco, San Francisco, CA 94117-1080}
\affiliation{Light Bridges S. L., Observatorio Astronómico del Teide. Carretera del Observatorio del Teide, s/n, Güímar, Santa Cruz de Tenerife, Canarias, Spain}
\email{osoler@usfca.edu}
\begin{abstract}


Multiply imaged, gravitationally lensed supernovae are rare but powerful tools for providing independent measurements on cosmological parameters. Supernova (SN) 2025wny (``SN~Winny") is the first gravitationally-lensed Type I superluminous supernova and the first lensed supernova in a galaxy-scale system that is suitable for time-delay cosmography studies. In this work, we present high-resolution $K_p$-band adaptive optics imaging of SN~Winny obtained with the near-infrared camera (NIRC2) on the W. M. Keck II telescope. With exquisite image quality (FWHM$\approx0\farcs065$) we determine and make use of the precise astrometric positions of the five multiple images as constraints for our lens mass models. With \texttt{lenstronomy} and {\sc Glee}, we parameterize the total mass of the system with a singular isothermal ellipsoid, a singular isothermal sphere, and external shear. The two independent models are in excellent agreement and reproduce the observed image positions with sub-milli-arcsecond residuals. The inferred projected total masses enclosed within the Einstein radii of the primary and secondary lens galaxies are M$_1$~=~4.44$^{+0.06}_{-0.05}\times10^{11}~M_\odot$ and M$_2$~=~0.96$^{+0.02}_{-0.02}\times10^{11}~M_\odot$, respectively. Likewise, the inferred effective velocity dispersion of the primary lens is $\sigma_{1} =$~277.4$^{+0.9}_{-0.7}$ km/s, consistent with the independent spectroscopic measurement made by DESI of $\sigma_{\star,1} =$~298$\,\pm\,37$ km/s. Our modeling results are also consistent with previous results for the same system
with data from the Large Binocular Telescope (LBT), using the same lens modeling codes. We also corroborate their finding that the SN multiple image A has an anomalous excess of flux by a factor of $\sim2-3$ beyond what our smooth mass models predict.
\end{abstract}

\setcounter{footnote}{1}
\section{Introduction}
Strong gravitational lensing is a powerful tool to study a variety of physical phenomena. In particular, strong lensing of supernovae (SNe) is emerging as a unique probe of cosmology through time-delay cosmography.  
By measuring the relative time delay between the multiple images of a lensed SN, it is possible to measure the value of the Hubble constant, $H_{0}$ \citep[e.g.,][]{kelly2023a,pierel2024,grayling2026a}. 
This has typically been done with strongly lensed quasars \citep[e.g.,][]{wong2020,tdcosmo2025}, which are, at this moment, more numerous and can be monitored continuously for years or decades \citep[e.g.,][]{millon2020b,dux2025}. However, with advancing observational capabilities and dedicated surveys, the sample of strongly lensed supernovae that are being discovered is increasing \citep[e.g.,][]{kelly2015b,rodney2021a}, with hundreds of such systems predicted to be discovered by Rubin/LSST \citep{abe2025}.

SN 2025wny (``Winny") was discovered in September 2025 by the Zwicky Transient Facility \citep{bellm2019a} and reported by \citet{taubenberger2025} and \citet{johansson2025}. 
The lens system which hosts SN~Winny was originally identified as a strong lens candidate by \citet{canameras2020a} as part of a convolutional neural network based search for strong lenses in the Panoramic Survey Telescope and Rapid Response System (Pan-STARRS) 3$\pi$ imaging survey.
Spectroscopic follow-up observations show that SN~Winny is a superluminous SN (SLSN) at $z =2.008\pm0.001$\footnote{\citet[][]{johansson2025} reports a host redshift of $z=$2.011$\pm0.001$, we use $z=$2.008 for consistency with \citet[][]{ecker2026}}, strongly lensed into multiple images by two foreground lens galaxies at $z = 0.375$.  Follow-up imaging from the Canada France Hawaii Telescope (CFHT) by \citet{aryan2025} and the Large Binocular Telescope (LBT) by 
\citet[][hereafter \citetalias{ecker2026}]{ecker2026}
have recently revealed the presence of five multiple images of SN~Winny. SN~Winny is the first strongly lensed SLSN, and only the third known galaxy-scale lensed SN. Of particular interest is the fact that SN~Winny is the first galaxy-scale lensed SN that could potentially be used for time-delay cosmography, as the two previous known systems \citep{goobar2017a,goobar2022a} have time delays too short to yield high-precision cosmological measurements. 

In this paper, we present new high-resolution $K_p$-band adaptive optics imaging data of SN~Winny using the Near-infrared Camera, Second Generation (NIRC2) on the Keck II telescope, along with preliminary independent lens models based on these data. We make use of two well-tested lens modeling algorithms, \texttt{lenstronomy} \citep{birrer2018a} and Gravitational Lens Efficient Explorer \citep[\textsc{Glee};][]{suyu2010b,suyu2012}.

The paper is structured as follows. Section~\ref{sec:data} describes the NIRC2 observations and data reduction process. In Section~\ref{sec:modeling}, we outline the modeling procedure, including the two algorithms used and the total mass parameterization. We present and discuss our modeling results in Section~\ref{sec:results} and compare them with those reported by E26 in Section~\ref{sec:ecker}. We summarize our conclusions in Section~\ref{sec:conclusions}.  Throughout this paper, we assume a flat $\Lambda$CDM cosmology with $H_{0}=70~\mathrm{km~s^{-1}~Mpc^{-1}}$ and $\Omega_{\mathrm{m}} = 0.3$.  

\section{Observations \& Data} \label{sec:data}
As an extraordinarily unique system, SN~Winny has been (and continues to be) the target of multiple photometric and spectroscopic campaigns. In particular, CFHT/MegaCam and LBT/LUCI observations have revealed a complex lensing system, where two lens galaxies (labeled with G1 and G2) strongly lens SN~Winny into five multiple images, identified with A-E (\citealp{aryan2025}; \citetalias{ecker2026}). DESI/DR1 spectroscopy provides a redshift measurement for G1, with $z=0.375$\footnote{\url{https://www.legacysurvey.org/viewer/desi-spectrum/dr1/targetid39633043073273740}} and a stellar velocity dispersion measurement\footnote{\url{https://data.desi.lbl.gov/doc/releases/dr1/vac/fastspecfit/}} of $\sigma_{\star, 1}=298 \pm  37 ~\rm km~s^{-1}$, obtained with the \texttt{fastspecfit} pipeline\footnote{\url{https://github.com/desihub/fastspecfit}}. The second deflector galaxy, G2, was spectroscopically confirmed to lie at the same redshift as G1 thanks to follow-up observations with the Nordic Optical Telescope \citep{taubenberger2025}.

Here, we focus on the newly obtained high-angular resolution Keck/NIRC2 observations of SN~Winny along with the data reduction methodology in Section \ref{sec:Keck/NIRC2}. The modeling of the surface-brightness distribution of the two lens galaxies and the five multiple images of SN~Winny is summarized in Section \ref{subsec:psf}.


\subsection{Keck/NIRC2} \label{sec:Keck/NIRC2}
On 11 November 2025, we triggered a Target of Opportunity (ToO; PI: B.~J.~Shappee) observation of SN~Winny using the NIRC2 near-infrared imager on Keck II with laser guide star (LGS) adaptive optics (AO). We used the smallest field of view configuration of 10$^{\prime\prime}\times$10$^{\prime\prime}$, which has a pixel scale of 0\farcs009952. Traditionally, NIRC2 uses a small-box or three-point dither pattern, where the target remains on the detector at each position for sky subtraction. Due to the size of the system ($\sim5^{\prime\prime}$ across), we dither completely off-target (``on-off dithers") while keeping the AO loop closed to obtain dedicated sky frames. Due to technical difficulties with the AO system, we obtained only one 180-second exposure on the science target and three 180 sec exposures of the sky background in the \textit{K$_p$} (2.12 $\mu$m) filter. We also took dedicated observations of the tip-tilt star ($\sim31^{\prime\prime}$ off-axis) on the NIRC2 science detector to construct an empirical point-spread function (PSF) model (described in Section~\ref{subsec:psf}). The tip-tilt star observations used a standard 3-point dither pattern with 60 seconds of exposure time at each position. We estimate the full width at half maximum (FWHM) of these observations to be $\sim$0\farcs065 using a Gaussian approximation for the PSF. We show the image of this system in Figure~\ref{fig:data} in the left panel. For visualization purposes, in the right panel we show the image convolved with a Gaussian kernel with a $\sigma$ width equivalent to the FWHM.

Reduction of these observations was performed using the \texttt{nirc2\_reduce} pipeline \citep{molter2022}, which was originally developed for solar system objects observed as a part the Keck Twilight Zone program. It natively supports reduction for observations using standard dither patterns as well as on-off dithers. For the science frame, we used the median of the three sky frames to construct a sky background and for the tip-tilt star, we used the empty portions of the detector to estimate and subtract the sky background. We apply standard dark subtraction, flat-fielding, and cosmic ray removal via \texttt{nirc2\_reduce}. The images were also corrected for geometric distortion of the NIRC2 detector according to the solution provided by \citet{lu2016}. In addition to distortions, there is also a difference between the orientation given in the header of the NIRC2 images and the true orientation which is measured to be 0.252$^{\circ}$ East of North relative to the true sky position angle \citep[][]{yelda2010}. We also correct for this effect. Figure~\ref{fig:data} shows the fully-reduced \textit{K$_p$} science frame with labels for the lens galaxies and SN multiple images. 

\vspace{0.25cm}
\begin{minipage}{0.99\textwidth}
\makebox[\textwidth][c]{
  \includegraphics[keepaspectratio=true,width=0.95\textwidth]{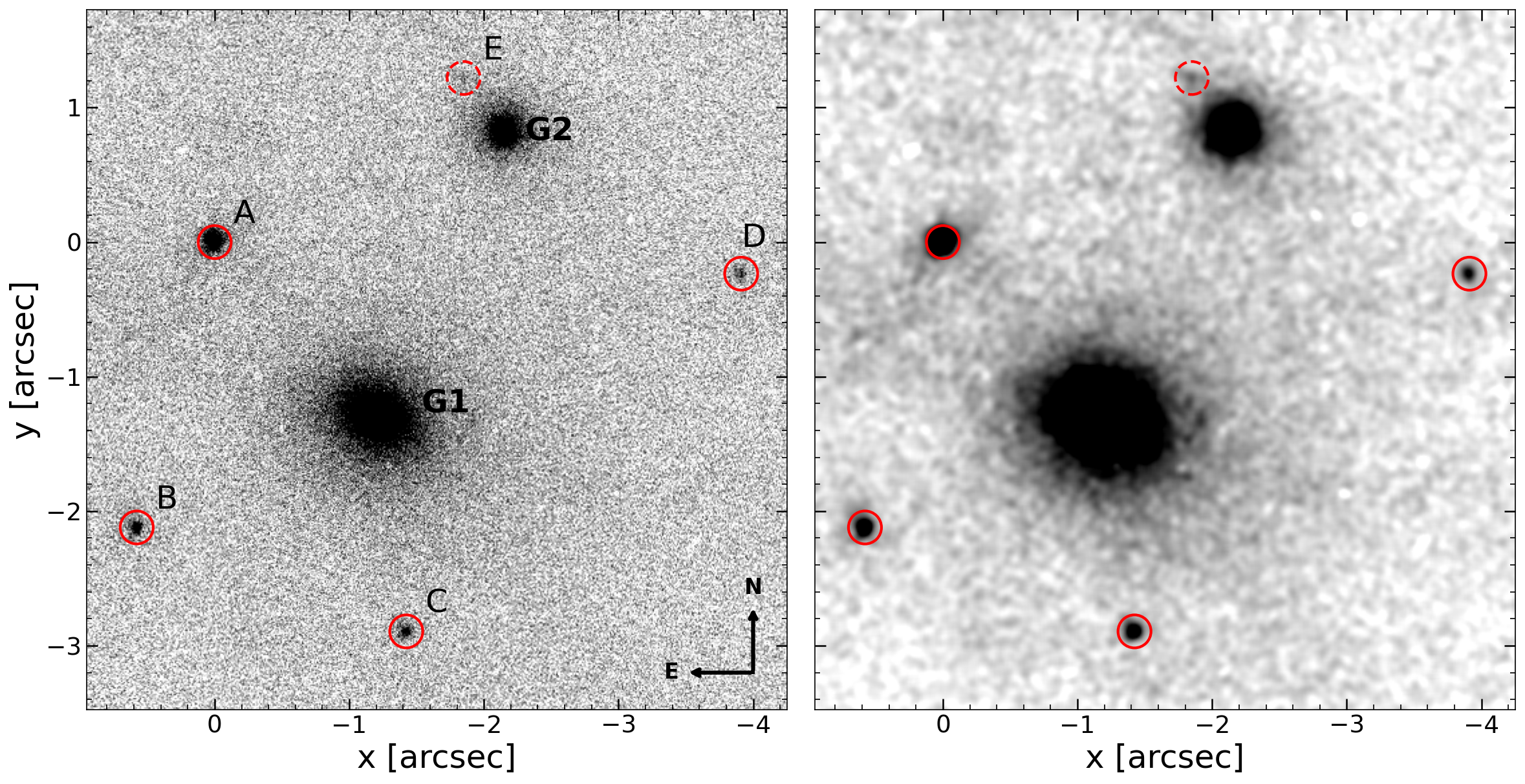}}
\captionof{figure}{Keck/NIRC2 imaging data of SN~Winny in the \textit{K$_p$} band. Following the convention introduced by previous works, we label the two lens galaxies with G1 and G2. The multiple images of SN~Winny are identified with A–E. Both images are $5.2\arcsec \times 5.2\arcsec$, North is up and East is left. \textit{Left}: The fully reduced image of SN~Winny with no alterations. \textit{Right}: The image of SN~Winny convolved with a Gaussian kernel with a $\sigma$ width equivalent to the FWHM (0\farcs065). This image also uses more dramatic clipping of pixel values.}
\label{fig:data}
\end{minipage}

\subsection{PSF \& Lens Light Model}
\label{subsec:psf}
Dedicated observations of the tip-tilt star allow us to produce an empirical PSF (ePSF). With the exception of the multiple image A, the SN multiple images do not have a signal-to-noise ratio which we deem to be sufficient to construct a reliable ePSF \citep[see][]{chen2016}. The tip-tilt star is $\sim$31$^{\prime\prime}$ off-axis from the science target and thus there may be differences in the shape of the PSF wings when comparing to the PSF wings of the SN multiple images. Despite this, we expect the core of the PSF to be stable. For the purposes of this work, which require only the determination of centroids, the use of the tip-tilt star for an ePSF is thus sufficient. 

In order to construct the ePSF, we used the \texttt{Point Spread Function reconstruction} \citep[\texttt{PSFr}; Birrer et al. in prep,][]{birrer2018a} algorithm. \texttt{PSFr} generally performs an iterative PSF reconstruction from multiple images of stars (or other point sources). Since we do not have images of multiple point sources (with consistent shapes), we employed \texttt{PSFr} on the individual, sky-background subtracted images of the tip-tilt star. \texttt{PSFr} accounts for sub-pixel astrometric shifts between the tip-tilt star images, where any sub-pixel shifts are applied via interpolation back onto the native pixel grid before combining. We do not over-sample the ePSF, and thus we do not recover the intra-pixel structure. While the difference between this ePSF model and a simple stack of the tip-tilt star images is minimal in practice, this method is more robust and avoids potential smearing of the PSF core. 

We simultaneously fit the light distribution of both lens galaxies and the five SN multiple images using {\sc Glee} \citep{suyu2010b,suyu2012}. The five SN multiple images are modeled with point sources convolved with the PSF, with their centroids and amplitudes left as free parameters. The lens galaxies are instead modeled as elliptical S\'{e}rsic profiles. Based on several modeling parameterizations, we find that G1 is best characterized by the sum of two S\'{e}rsic profiles with a common centroid, while G2 is adequately characterized by a single S\'{e}rsic profile. All S\'{e}rsic profiles are given flat priors on the values of the minor-to-major axis ratio, $0.3 \leq q \leq 1.0$, effective radius, $0\arcsec \leq r_{\mathrm{eff}} \leq 10\arcsec$, and S\'{e}rsic index, $0.4 \leq n \leq 10.0$. The centroid and position angle (defined East of North) are free parameters (unbounded, uniform priors), while the amplitude must be non-negative, but can otherwise vary freely.  Other than the common centroid, the two S\'{e}rsic profiles for G1 are allowed to vary independently.  The fitting is performed using a Markov chain Monte Carlo (MCMC) sampler.  The best-fit light model and the normalized residuals are presented in Figure~\ref{fig:light}, showing the goodness of our fits. The marginal posterior median values, with the associated $1\sigma$ uncertainties, of the positions and fluxes from the light models of the seven objects are reported in Table~\ref{tab:positions}. 

\subsection{Multiple image E}
\label{subsec:img_E}
The initial detection of the fifth multiple image, E, was reported by \citet[][]{aryan2025} thanks to deep imaging from the Canada-France-Hawaii Telescope (CFHT). While the detection of multiple image E at the time of discovery was tentative, the detection and location of the multiple image E was confirmed by \citetalias{ecker2026}. The multiple image E is located $\sim0\farcs2$ North and East of the secondary lens galaxy, G2. In the Keck NIRC2 imaging presented here, image E is detected at the $\sim7.5\sigma$ level according to the flux uncertainty reported from the model fit. This detection was made by the {\sc Glee} light model with a flat, open prior on the position and was directly included in the model. The right panel of Figure~\ref{fig:data} accentuates this detection. 

\vspace{0.25cm}
\begin{minipage}{0.99\textwidth}
\makebox[\textwidth][c]{
  \includegraphics[keepaspectratio=true,width=0.99\textwidth]{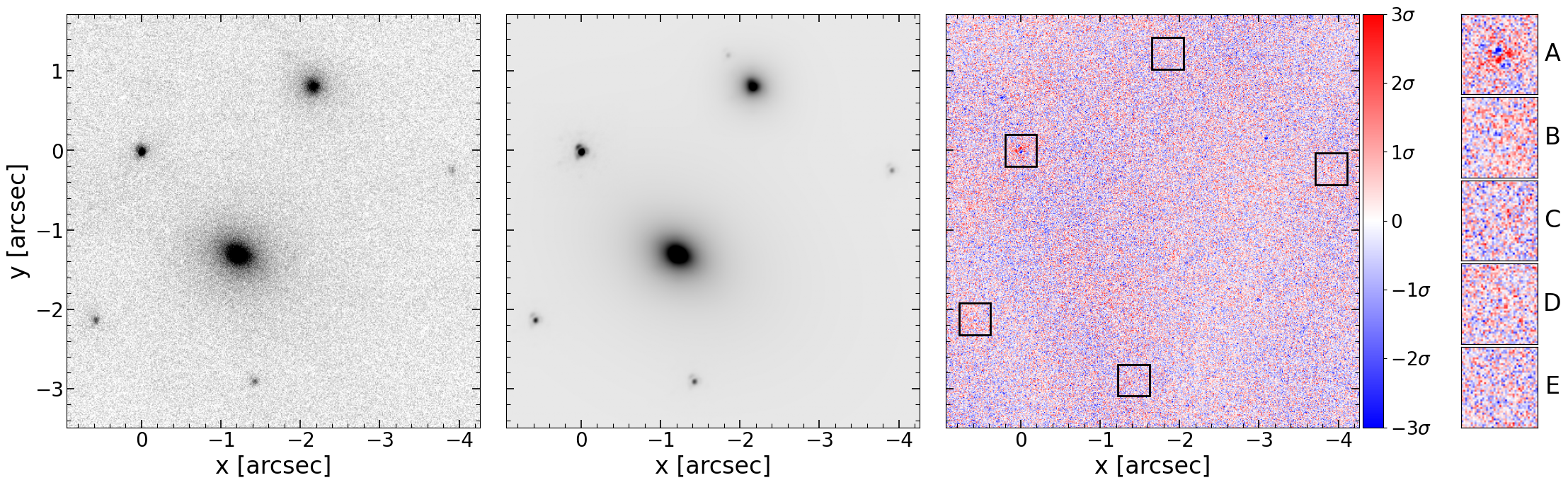}}
\captionof{figure}{Keck/NIRC2 imaging data (left panel), the best-fit light model from {\sc Glee} (middle panel), and the normalized residual image within a range between $-3 \sigma$ and $3 \sigma$ with boxes centered on the multiple images A$-$E (right panel). The primary panels are $5.2\arcsec \times 5.2\arcsec$. North is up and East is left.  Shown on the right-most end of the figure, are $0.4\arcsec \times 0.4\arcsec$ zoom-in cutouts of the normalized-residual image, centered on each of the multiple images.}
\label{fig:light}
\end{minipage}

\begin{table}[H]
\centering
\caption{Astrometric positions for the two lens galaxies and five SN multiple images measured relative to that of the multiple image A. The estimated flux values for the SN multiple images are provided in arbitrary units.}
\label{tab:positions}
\begin{tabular*}{0.75\textwidth}{c @{\extracolsep{\fill}} r r r r}
\hline\hline
\multicolumn{1}{c}{Object} & \multicolumn{1}{c}{$x$ (\arcsec)} & \multicolumn{1}{c}{$y$ (\arcsec)} & \multicolumn{1}{c}{Flux (arbitrary units)} \\
\hline
G1 & $-1.2167^{+0.0005}_{-0.0005}$ & $-1.3021^{+0.0004}_{-0.0004}$ & \multicolumn{1}{c}{---}\\
G2 & $-2.1624^{+0.0006}_{-0.0006}$ & $0.8217^{+0.0006}_{-0.0007}$ & \multicolumn{1}{c}{---}\\
A  & $0.0000^{+0.0003}_{-0.0003}$ & $0.0000^{+0.0003}_{-0.0003}$ & \multicolumn{1}{c}{\makebox[5em][r]{$159.9^{+1.1}_{-1.1}$}\hspace{1em}} \\
B  & $0.5791^{+0.0011}_{-0.0023}$ & $-2.1210^{+0.0009}_{-0.0009}$ & \multicolumn{1}{c}{\makebox[5em][r]{$ 44.8^{+1.0}_{-1.0}$}\hspace{1em}} \\
C  & $-1.4232^{+0.0012}_{-0.0012}$ & $-2.8937^{+0.0020}_{-0.0013}$ & \multicolumn{1}{c}{\makebox[5em][r]{$ 34.2 ^{+1.0}_{-1.0}$}\hspace{1em}} \\
D  & $-3.9113^{+0.0017}_{-0.0019}$ & $-0.2348^{+0.0030}_{-0.0035}$ & \multicolumn{1}{c}{\makebox[5em][r]{$ 21.8 ^{+1.0}_{-1.0}$}\hspace{1em}} \\
E  & $-1.849^{+0.005}_{-0.006}$ & $1.220^{+0.006}_{-0.007}$ & \multicolumn{1}{c}{\makebox[5em][r]{$  7.8^{+1.0}_{-1.1}$}\hspace{1em}} \\
\hline
\end{tabular*}
\end{table}


\section{Lens modeling methodology}
\label{sec:modeling}
In this work, we make use of two well-established, independent, lens modeling algorithms, \texttt{lenstronomy} \citep[][]{birrer2018a} and \textsc{Glee} \citep{suyu2010b,suyu2012}. While these two modeling algorithms take similar approaches in parameter estimation and sampling, the exact methodology and optimization choices may vary between models. In the following section, we discuss the commonalities between approaches and also the relevant differences. We also discuss the parameterization of the total mass distribution of the system and the associated priors.

\subsection{Mass Parameterization \& Priors}
\label{subsec:priors}

Choices for the total mass parameterization and priors were implemented as uniformly as possible between the two modeling algorithms. We parameterize G1 with a singular isothermal ellipsoid (SIE) and G2 with a singular isothermal sphere (SIS) profiles. The light profile of G2 indicates that it is more circular than G1. We also include an external shear component, $\gamma_{\mathrm{ext}}$, to account for perturbations due to line-of-sight structure and model complexity. We consider these choices to provide a model with the minimum complexity required to accurately describe the system while also minimizing the number of free parameters. 

The five SN multiple image positions are used as observables in our lens models. The adopted positions are taken to be the posterior median values from the light fitting procedure described in Section~\ref{subsec:psf}.  Although the statistical uncertainties on the positions of images A$-$D can be on the order of a milli-arcsecond or less, we assume an uncertainty of 0\farcs004 in both the $x$ and $y$ directions to account for deflection by undetected substructure in the lens galaxies or along the line of sight \citep{chen2007,takahashi2014,birrer2025b}.  We assume a larger uncertainty in the centroid of multiple image E of 0\farcs01. This is estimated considering the FWHM (0\farcs065) of the observations and the signal-to-noise ratio of image E (7.5), assuming a Gaussian PSF.
These positions provide ten observables (two coordinates per multiple image) with a total of twelve free parameters. These parameters include five to describe G1, three for G2, two for the external shear, and an additional two nuisance parameters for the unlensed source position. The system is therefore not fully constrained by the positions of the point-like multiple images alone, and we rely on well-informed priors for the mass centroids of G1 and G2, guided by the observed light centroids, in our modeling procedure. This choice is well motivated, as strong lensing analyses of galaxy-scale lenses have shown that mass and light centroids are generally consistent \citep[e.g.][]{bolton2008a}. 
We show the full set of priors for each of the model parameters in Table~\ref{tab:priors}. These priors are effectively identical in the two modeling approaches. Differences in parameter definitions may result in slightly different bounds for uniform priors, including the Einstein radius, ellipticity, and external shear. 


\begin{table}[H]
\centering
\caption{Prior distributions for the lens model parameters.}
\label{tab:priors}
\renewcommand{\arraystretch}{1.2}
\begin{threeparttable}
\begin{tabular}{llcl}
\hline\hline
Component & Parameter & Prior & Description \\
\hline
G1 Mass (SIE) & $\theta_{\mathrm{E,1}}$ & $\mathcal{U}(0.5,\, 2.5)$ & Einstein radius [$\arcsec$] \\
        & $x_{1}$                 & $\mathcal{N}(x_\mathrm{G1},\, 0.05)$ & Mass $x$-center [$\arcsec$] \\
        & $y_{1}$                 & $\mathcal{N}(y_\mathrm{G1},\, 0.05)$ & Mass $y$-center [$\arcsec$] \\
        & $q_1$                   & $\mathcal{U}(0.3,\, 1.0)$ & Minor-to-major axis ratio \\
        & $\varphi_1$             & $\mathcal{U}(0,\inf)$ & Position angle [$^{\circ}$] \\
\hline
G2 Mass (SIS) & $\theta_{\mathrm{E,2}}$ & $\mathcal{U}(0.1,\, 1.5)$ & Einstein radius [$\arcsec$] \\
        & $x_{2}$                 & $\mathcal{N}(x_\mathrm{G2},\, 0.05)$ & Mass $x$-center [$\arcsec$] \\
        & $y_{2}$                 & $\mathcal{N}(y_\mathrm{G2},\, 0.05)$ & Mass $y$-center [$\arcsec$] \\
\hline
External shear & $\gamma_{\mathrm{ext}}$  & $\mathcal{U}(0.0,\, 0.5)$ & Magnitude \\
               & $\varphi_{\mathrm{ext}}$ & $\mathcal{U}(0,\inf)$ & Position angle [$^{\circ}$] \\
\hline
\end{tabular}
\begin{tablenotes}
\footnotesize
\item \textbf{Note.} $\mathcal{U}(a,b)$ denotes a uniform prior over $[a,b]$ and $\mathcal{N}(\mu,\sigma)$ a Gaussian prior with mean $\mu$ and standard deviation $\sigma$. $x_\mathrm{G1}$, $y_\mathrm{G1}$, $x_\mathrm{G2}$, $y_\mathrm{G2}$ denote the lens light centers for G1 and G2 (see Table~\ref{tab:positions}). 
\end{tablenotes}
\end{threeparttable}
\end{table}

\subsection{Modeling Procedure}
\label{subsec:lenstronomy}
The first modeling algorithm employed is the public software \texttt{lenstronomy} \citep{birrer2018a}. As opposed to other modeling pipelines, such as {\sc Glee}, \texttt{lenstronomy} defines the ellipticity and external shear in Cartesian components $(e_1, e_2)$ and $(\gamma_1, \gamma_2)$, respectively, rather than the axis ratio ($q$) and position angle ($\varphi$). Since a uniform prior on $(q, \varphi)$ is not equivalent to a uniform prior on the Cartesian components, we enforce the priors listed in Table~\ref{tab:priors} through a custom log-likelihood that rejects any sample with $q < 0.3$, $q > 1$, or $\gamma_{\rm ext} > 0.5$, where the axis ratio and shear magnitude are recovered from the Cartesian components at each evaluation. This ensures that priors are as consistent as possible between each modeling algorithm.

The five observed multiple image positions are treated as fixed constraints in \texttt{lenstronomy}'s point source modeling. For a trial set of lens parameters, each image position $\boldsymbol{\theta}_i$ is ray-traced to the source plane via the lens equation, $\boldsymbol{\beta}_i = \boldsymbol{\theta}_i - \boldsymbol{\alpha}(\boldsymbol{\theta}_i)$. The likelihood penalizes the scatter among the resulting source positions $\{\boldsymbol{\beta}_i\}$, with the image-plane astrometric uncertainty propagated to the source plane at each image position. 
This approach evaluates the goodness of fit on the source plane rather than on the image plane. However, because the Hessian propagation accounts for the local magnification at each image position, images near critical curves are appropriately up-weighted, and the two approaches yield consistent results for well-fitting models with small source-plane residuals \citep{birrer2018a}. We note that {\sc Glee} evaluates its likelihood directly on the image plane. This choice of evaluation plane is a methodological difference between the two codes.

For certain parameter combinations, this parameterization of the system total mass distribution can produce seven multiple images of SN~Winny rather than the five that are observed. 
In order to enforce the observed image multiplicity, we impose an additional constraint. At each likelihood evaluation in which the lens equation is solved, a log-likelihood penalty is applied if the additional images (beyond the five observed) are predicted.
This effectively excludes regions of the parameter space where the critical curve topology permits extra image pairs. This, in turn, manifests as a sharp boundary in the posterior, in particular for the ellipticity of G1, which is influential on the image multiplicity for this lens geometry.
Parameter estimation occurs in three stages within \texttt{lenstronomy}'s \texttt{FittingSequence} framework. Particle swarm optimization (PSO) performs the initial global search followed by Nelder--Mead simplex to refine the PSO solution. Posterior distributions are sampled using the MCMC ensemble sampler \texttt{emcee}.

The modeling procedure used with {\sc Glee} takes a more straightforward approach. {\sc Glee} directly samples the posterior probability distribution using an MCMC algorithm.  After an initial MCMC chain, a covariance matrix is generated, and this matrix is used along with the current parameter values to generate a proposal distribution for the subsequent chain.  The likelihood is evaluated on the image plane based on the constraints described in Section~\ref{subsec:priors}.
A flat prior on the magnitude and position angle of these quantities is not equivalent to a flat prior on the Cartesian components, but the effect is likely minimal as seen in the posteriors in Figure~\ref{fig:posterior}.


\section{Modeling Results}
\label{sec:results}
In this section, we report and compare the modeling results from the two independent lens modeling algorithms, \texttt{lenstronomy} and {\sc Glee}. We report the median value from the marginalized posterior of each parameter. Quoted uncertainties denote the 16th and 84th percentiles of the posterior distributions, equivalent to the $\pm1\sigma$ interval for a Gaussian distribution. This choice is robust to non-Gaussianity and accommodates asymmetric posteriors. Generally, each lens modeling method produces consistent values of predicted lens mass parameters, and the two lens models reconstruct the observed multiple image positions to less than the assumed uncertainty of $0\farcs004$. We further evaluate the consistency of the modeling codes quantitatively.


In order to enable a direct comparison, we first convert the relevant parameters such that they are defined consistently between the two analyses. 
All values for the Einstein radius ($\theta_{\mathrm{E}}$) use the spherical-equivalent approximation.
The Cartesian ellipticity and shear components $(e_1, e_2)$ and $(\gamma_1, \gamma_2)$ used internally by \texttt{lenstronomy} are converted to the equivalent axis ratio and position angle $(q_1, \varphi_1)$ and shear magnitude and angle $(\gamma_{\mathrm{ext}}, \varphi_{\mathrm{ext}})$. Following these conversions, all the median values with the associated 16th and 84th percentile uncertainties are reported in Table~\ref{tab:model_params} and the respective posterior distributions are shown in Figure~\ref{fig:posterior}.

With consistent definitions across model parameters, we proceed with a direct statistical comparison. We quantify the agreement between the two lens models using the same standard pair-wise comparison as in \citetalias{ecker2026}:

\begin{equation}
\Delta_{i,j} = \frac{|\theta_i - \theta_j|}{\sqrt{\sigma_i^2 + \sigma_j^2}}, i \neq j \, ,
\label{eq:consistency}
\end{equation}

\noindent where $\Delta_{ij}$ represents the difference in units of standard deviations, $\theta_i$ and $\theta_j$ are the posterior median values from each modeling code, normalized by their combined uncertainties $\sigma_i$ and $\sigma_j$, which, in this case, are the symmetrized uncertainties derived from the 16th and 84th percentiles. 
The comparison across all ten model parameters shows a remarkable agreement, with no parameter values exceeding a difference of $0.6\sigma$ and an average difference of 0.3$\sigma$ (see Table~\ref{tab:model_params}). 
The Einstein radii of G1 and G2 are among the most tightly constrained parameters in both models, they both yield $\theta_{\mathrm{E,1}} \sim 1.58\arcsec$ for G1, and $\theta_{\mathrm{E,2}} \sim 0.74\arcsec$.

\begin{table}[h]
\centering
\caption{Median posterior values and corresponding 1$\sigma$ uncertainties of lens model parameters from the two independent modeling codes.}
\label{tab:model_params}
\renewcommand{\arraystretch}{1.2}
\begin{threeparttable}
\begin{tabular*}{0.7\textwidth}{l c @{\extracolsep{\fill}} r r c}
\hline\hline
 & & \multicolumn{2}{c}{Best-fit value} \\
\cline{3-4}
\multicolumn{1}{c}{Component} & \multicolumn{1}{c}{Parameter [Unit]} & \multicolumn{1}{c}{\texttt{lenstronomy}} & \multicolumn{1}{c}{\sc Glee} & $\Delta$ [$\sigma$]\\
\hline
G1 Mass (SIE) & $\theta_{\mathrm{E,1}}$ [\arcsec] & $  1.587^{+0.010}_{-0.008}$   & $  1.581^{+0.013}_{-0.013}$& 0.38\\
        & $x_{1}$ [\arcsec]                 & $  -1.143^{+0.010}_{-0.010}$   & $  -1.148^{+0.010}_{-0.010}$& 0.35\\
        & $y_{1}$ [\arcsec]                 & $ -1.388^{+0.001}_{-0.001}$   & $ -1.388^{+0.004}_{-0.005}$& 0.00\\
        & $q_1$                   & $  0.73^{+0.02}_{-0.03}$   & $  0.75^{+0.02}_{-0.03}$& 0.44\\
        & $\varphi_1$ [$^{\circ}$]             & $ 50.66^{+6.13}_{-6.17}$   & $ 55.19^{+7.14}_{-6.81}$& 0.49\\
\hline
G2 Mass (SIS) & $\theta_{\mathrm{E,2}}$ [\arcsec]& $  0.736^{+0.008}_{-0.007}$  & $  0.734^{+0.010}_{-0.010}$& 0.16\\
        & $x_{2}$   [\arcsec]       & $  -2.180 ^{+0.039}_{-0.038}$  & $  -2.181^{+0.041}_{-0.037}$& 0.02\\
        & $y_{2}$   [\arcsec]       & $  0.872^{+0.032}_{-0.028}$  & $  0.849^{+0.039}_{-0.041}$& 0.46\\
\hline
External shear & $\gamma_{\mathrm{ext}}$  & $  0.11^{+0.01}_{-0.01}$   & $  0.12^{+0.01}_{-0.01}$& 0.57\\
               & $\varphi_{\mathrm{ext}}$ [$^{\circ}$]  & $-64.88^{+2.55}_{-2.33}$   & $-65.41^{+2.43}_{-2.18}$& 0.16\\
\hline
\end{tabular*}
\begin{tablenotes}
\footnotesize
\item \textbf{Note.} Position angles for ellipticity ($\varphi_1$) and shear ($\varphi_{\mathrm{ext}}$) are measured East of North.
\end{tablenotes}
\end{threeparttable}
\end{table}

 \vspace{0.cm}
\begin{minipage}{0.99\textwidth}
\makebox[\textwidth][c]{
  \includegraphics[keepaspectratio=true,width=0.9\textwidth]{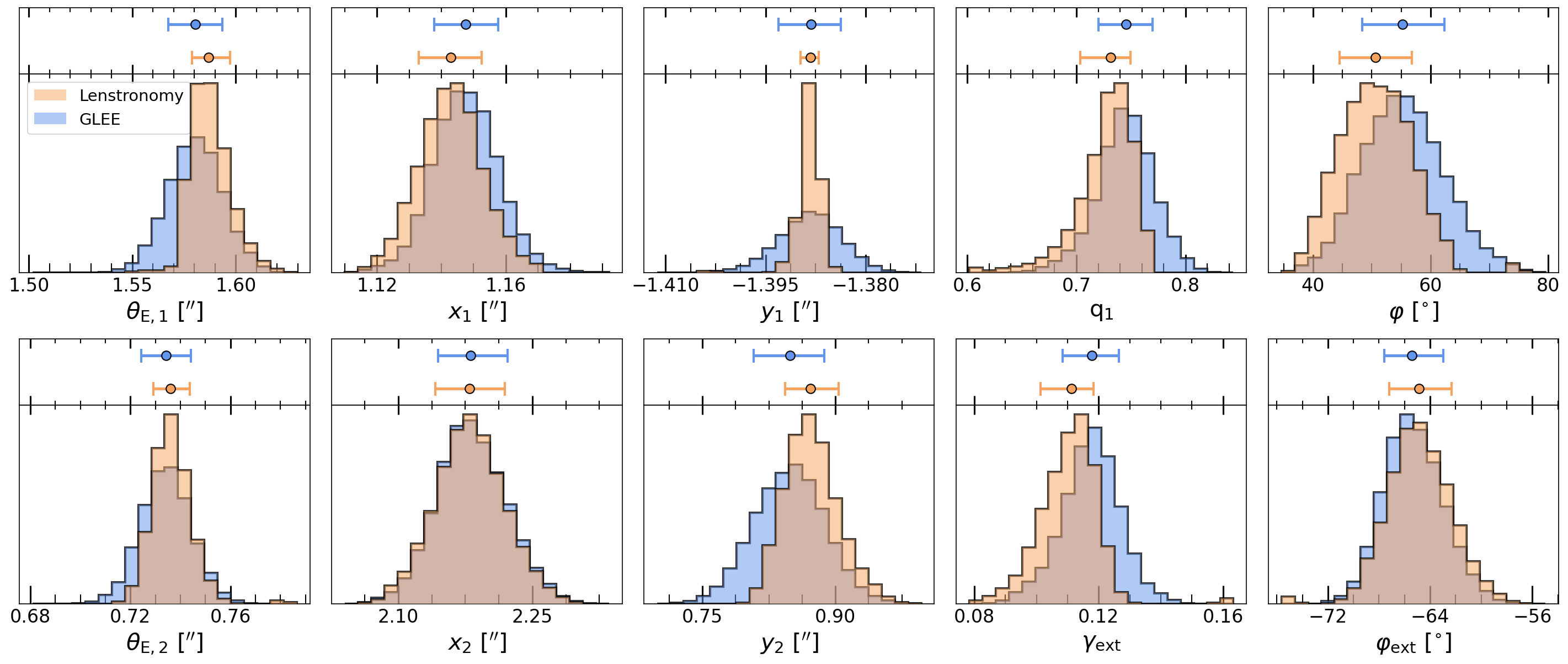}}
\captionof{figure}{Posterior distributions for the ten free parameters used in the modeling procedure for \threecodes. Above each posterior distribution, we show the median value with the 16th and 84th percentile range. See Table~\ref{tab:priors} for the descriptions of the parameters.} 
\label{fig:posterior}
\vspace{0.25cm}

\end{minipage}

 \vspace{0.5cm}
\begin{minipage}{0.99\textwidth}
\makebox[\textwidth][c]{
  \includegraphics[keepaspectratio=true,width=0.95\textwidth]{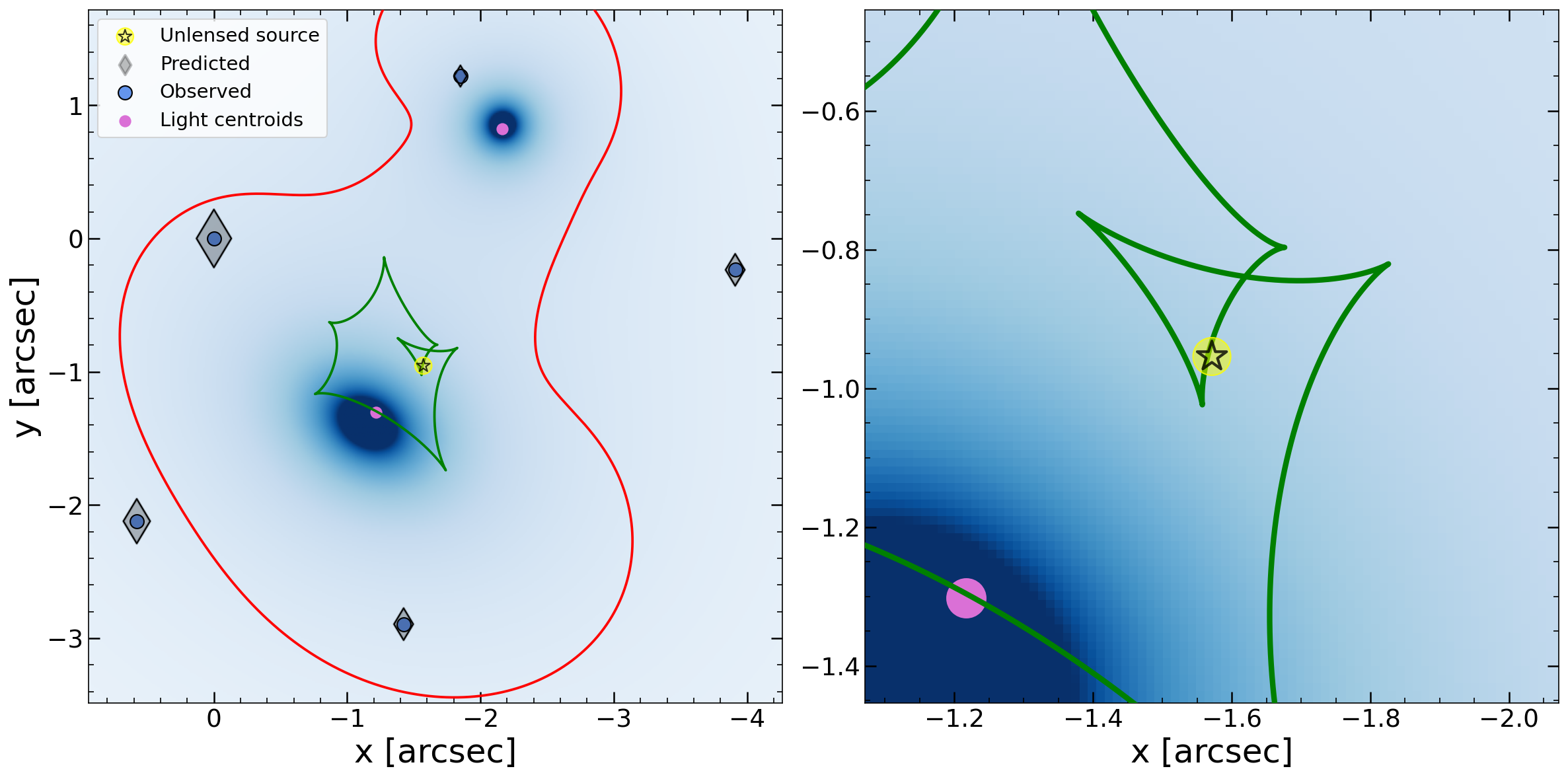}}
\captionof{figure}{Left: Convergence map of the final \texttt{lenstronomy} model. The caustic (green line) and unlensed-source position (yellow star) are shown on the source-plane and the critical curve (red line), predicted (grey diamonds), and observed image positions (blue points) are shown on the lens-plane along with the light centroids (pink points) of G1 and G2. The marker size for the predicted multiple image positions is scaled with the predicted magnifications reported in Table~\ref{tab:mags}. Right: A zoomed-in view of the caustic shape and the source position relative to the caustic edges.}
\label{fig:caustic}
\end{minipage}
\vspace{0.0cm}

In our models, the choice of total mass parameterization (SIE+SIS+shear) and their configurations can produce a caustic topology with an ``inner caustic area'', as shown in the right panel of Figure~\ref{fig:caustic}. This feature arises from the influence of G2 and separates regions of different image multiplicity on the source plane, in this case, either five or seven multiple images\footnote{The five and seven image configurations here do not include the two de-magnified, ``on-core" images}.
Other iterations of the \texttt{lenstronomy} model (without a restriction on image multiplicity) reveal a strong degeneracy in the source position and whether it falls interior or exterior to the inner caustic region. In all of these scenarios, the observed multiple image positions are recovered well within the 0\farcs004 uncertainty. However, models that predict seven images place the additional pair to the North-West of image A, with comparable magnifications. If these additional images were present, they would be detected in existing follow-up imaging. Furthermore, Figure~1 in \citetalias{ecker2026} shows an extended light emission of the SN host galaxy to the North-West of image A, implying that some portion of the host galaxy lies just within the inner caustic area, but not SN~Winny itself.
We therefore restrict our \texttt{lenstronomy} models to those that only predict the five observed multiple images (see Section~\ref{subsec:lenstronomy}). This restriction is reflected in the shape of the posterior distributions for some parameters shown in Figure~\ref{fig:posterior}. All the parameter posteriors for the {\sc Glee} model are roughly Gaussian, whereas \texttt{lenstronomy} shows noticeable non-Gaussianity for $\theta_{\mathrm{E,1}}$, $q_1$, and $\gamma_{\mathrm{ext}}$, each featuring sharp cut-offs (at low values for $\theta_{\mathrm{E,1}}$ and high values for $q_1$, and $\gamma_{\mathrm{ext}}$).
These boundaries are not a consequence of the priors, but rather reflect the image multiplicity constraint implemented in the \texttt{lenstronomy} model (see Section~\ref{subsec:lenstronomy} for details).
Both models from \threecodes~predict very similar caustic shapes and the same source position relative to the inner caustic area and thus reproduce the observables very well. 
The predicted caustics from GLEE are consistent, modulo a constant deflection that is degenerate with a translation in the source plane and has no effect on any observables.
Importantly, both models predict very similar source positions relative to the inner caustic.

Using the spherical definition for $\theta_{\mathrm{E}}$, combined with the lens and source redshifts, we can determine the corresponding effective velocity dispersion \citep[$\sigma$, a quantity that should be very similar to the central stellar velocity dispersion, $\sigma_{\star}$, see][]{Grillo2008} and the total mass enclosed ($M_{\mathrm{enc}}$) within the Einstein radii for G1 and G2. Since these properties are inferred directly from $\theta_{\mathrm{E}}$, and because $\theta_{\mathrm{E}}$ is consistent between the two modeling codes, we report the corresponding physical properties according to the \texttt{lenstronomy} median value for $\theta_{\mathrm{E}}$. For simplicity, we only propagate the statistical uncertainties associated with  $\theta_{\mathrm{E}}$. It is therefore likely that uncertainties quoted for $\sigma$ and $M_{\mathrm{enc}}$ are underestimated. The inferred velocity dispersion of G1 is $\sigma_{1} =$~277.4$^{+0.9}_{-0.7}$ km/s. This is consistent with the measured stellar velocity dispersion from the DESI spectrum of G1, reported by \texttt{fastspecfit} within the uncertainties. Our lens models also infer an effective velocity dispersion for G2 of $\sigma_{2} =$~188.7$^{+1.0}_{-0.9}$ km/s. Additional spectroscopic observations, which can provide a measurement of the stellar kinematics of the secondary lens galaxy, will be helpful to further constrain the lensing potential of the system. We also estimate the total mass enclosed within the Einstein radii of G1 and G2 to be M$_1$~=~4.44$^{+0.06}_{-0.05}\times10^{11}~M_\odot$ and M$_2$~=~0.96$^{+0.02}_{-0.02}\times10^{11}~M_\odot$, respectively.
The recovered mass centroids for both G1 and G2 lie within the $2\sigma$ range of the assumed prior distribution (0\farcs05 for \texttt{lenstronomy} and {\sc Glee}), which is centered on the light centroids reported in Table~\ref{tab:priors}. While the posterior distributions for the mass centroids do not show any irregularities that may be associated with a restrictive prior, the lack of sufficient degrees of freedom in our model limits our ability to state whether or not the mass centroids are inconsistent with the light centroids of G1 and G2.



\subsection{Magnifications \& Flux Ratios}
\label{subsubsec:flux_ratios}

The predicted magnification values, derived from the posterior distributions of the model parameters at the locations of the predicted multiple image positions,  from \threecodes~are reported in Table~\ref{tab:mags}. The reported values are the medians of magnifications for each chain and the uncertainties are the 16th and 84th percentiles. Due to the proximity of the source to the caustic (which formally denotes a line of infinite magnification), the predicted magnifications are highly sensitive to the precise location of the source relative to the caustic.
Despite this, all models recover the same image parity for all five multiple images. Images A, C, and E have negative magnifications, corresponding to saddle-points in the Fermat potential, while images B and D have positive magnification, corresponding to minima. Lensing theory would indicate that the total number of images is $N_{\mathrm{img}} = (2\times N_{\mathrm{saddle}})+1$ \citep{burke1981}. As mentioned previously, we do not include the additional two de-magnified ``on-core" images in our analysis. These images correspond to maxima in the Fermat potential. 
We apply the same comparison metric defined in Equation~\ref{eq:consistency} to the two sets of model-predicted magnification values and find $N_\sigma < 0.7$ for all five multiple images.

\begin{table}[H]
\centering
\caption{Magnification estimates for the five SN multiple images from \threecodes.}
\label{tab:mags}
\begin{threeparttable}
\begin{tabular*}{0.5\textwidth}{c @{\extracolsep{\fill}} r r}
\hline\hline
 & \multicolumn{2}{c}{Magnification ($\mu$)} \\
\cline{2-3}
\multicolumn{1}{c}{Image} & \multicolumn{1}{c}{\texttt{lenstronomy}} & \multicolumn{1}{c}{\textsc{Glee}} \\
\hline
A & $-8.8^{+1.9}_{-1.6}$ & $-10.3 ^{+2.1}_{-3.4}$\\
B & $6.1^{+0.4}_{-0.4}$ & $ 6.5 ^{+0.5}_{-0.5}$ \\
C & $-3.1^{+0.2}_{-0.2}$ & $-3.3 ^{+0.3}_{-0.3}$ \\
D & $3.3^{+0.1}_{-0.1}$ & $ 3.4 ^{+0.2}_{-0.1}$ \\
E & $-1.6^{+0.3}_{-0.3}$ & $-1.6 ^{+0.3}_{-0.3}$ \\
\hline
\end{tabular*}
\begin{tablenotes}
\footnotesize
\item \textbf{Note.} The sign of the magnification indicates image parity, whereas the flux amplification is given by $|\mu|$.
\end{tablenotes}
\end{threeparttable}
\end{table}

Following the methodology of \citetalias{ecker2026}, we compare the ratio of flux (in arbitrary units; Table~\ref{tab:positions}) to the model-predicted magnifications (Table~\ref{tab:mags}). Following the same convention and format of Figure~8 in \citetalias{ecker2026}, we show the comparison of the \texttt{lenstronomy} and {\sc Glee} flux-magnification ratios normalized by image B in Figure~\ref{fig:flux-mag}. The decision to normalize to image B is motivated by the lack of contamination by the host light near that multiple image. However, we find this decision is also physically motivated, placing the multiple images at minima of the Fermat potential (images B and D) at unity. The core findings between \citetalias{ecker2026} and this work are consistent; image A is anomalously bright by a factor of $\sim2-3$ in our smooth lens mass models. Images B and D are both within $1\sigma$ of unity for both models. While the flux excess of image A is the most dramatic, the multiple images C and E both deviate from unity by a noticeable amount. It is impossible to determine the cause of this anomalous flux with the available data. As identified in \citetalias{ecker2026}, the likely causes are one or more of (1) a lack of complexity in the smooth mass model, (2) micro-lensing by point-masses, (3) milli-lensing by substructure in the lens or along the line of sight, and (4) time-delay induced phase differences in the SN. We reasonably exclude host-light contamination in the photometry as a source of significant uncertainty due to the lack of any host light detection in the Keck NIRC2 observations (unlike in the LBT data).

 \vspace{0.5cm}
\begin{minipage}{0.99\textwidth}
\makebox[\textwidth][c]{
  \includegraphics[keepaspectratio=true,width=0.75\textwidth]{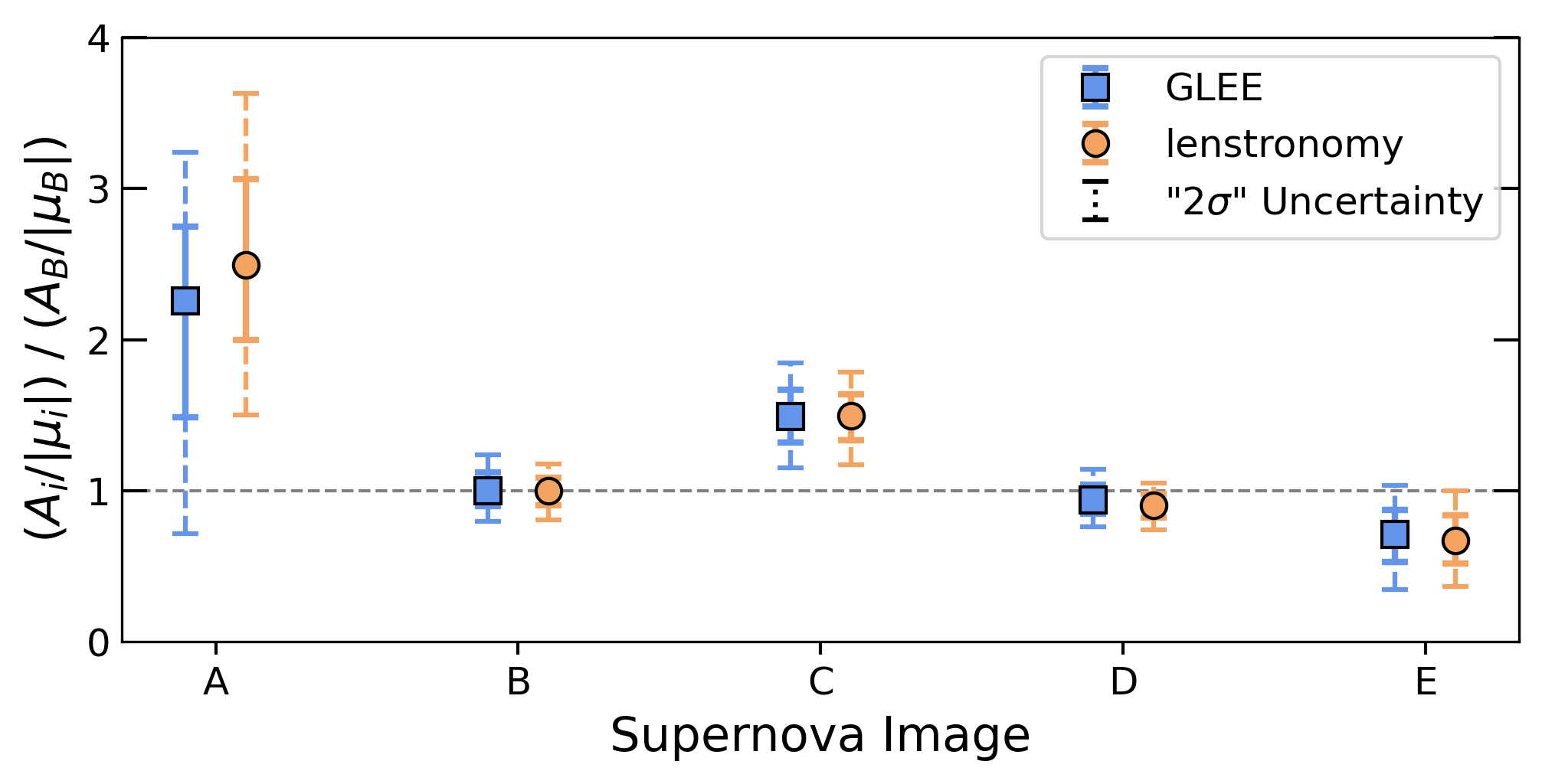}}
\captionof{figure}{Flux-to-magnification ratios for each of the five lensed images, normalized to image B with a horizontal dashed line at unity (1). We show the ratios for both the \threecodes~model predicted magnifications with 1 and $2\sigma$ intervals for each.}
\label{fig:flux-mag}
\end{minipage}
\vspace{0.25cm}


While we do not report time-delays in this work, it is worth noting that images A and C (and E) are roughly in-phase with relative time-delays between them on the order of a few days. Thus, in the rest-frame, the time-delays between these images are minimal compared to the expected timescale for significant evolution of the SN itself. Using the $r$-band light curve from \citet[][]{johansson2025}, we are able to approximate the day-to-day change in magnitude, with a max $\sim$0.025 mag/day in the observer frame, post peak. In \citet[][]{taubenberger2025}, magnitudes of A and C are reported to be 19.6 and 21.5, respectively, in the $r$-band observations from the Lulin One-meter Telescope (LOT). After correcting for lensing magnification, each image would be $21.95$ mag
and $22.83$ mag,
a difference of 0.92 mags. While this is not definitive, this supplies additional evidence that would suggest there is a real, astrophysical flux anomaly associated with at least image A but also likely image C, beyond a simple SN phase difference. This anomaly requires additional investigation.

We note that the magnification of image A, as predicted by the smooth lens mass model, is extremely sensitive to the model parameters due to its proximity to the critical curve. This is evidenced by the large fractional uncertainties shown in Table~\ref{tab:mags}. Being a saddle-point image, image A may be more sensitive to perturbations from intervening masses in the form of point masses or extended, low-mass dark matter subhalos \citep[micro- and milli-lensing][]{schechter2002a,dobler2006a,inoue2016}.
However, the observed flux excess in image A is in tension with the de-magnification bias that is expected in these instances. Nevertheless, individual perturbers can produce additional magnifications of either sign at a lensed image position. 

\section{Comparison with Ecker et al. 2026}
\label{sec:ecker}
In this section, we compare the results of our analysis to those of \citetalias{ecker2026}, which modeled the total mass distribution of the system based on data from the LUCI instrument on the LBT using \texttt{lenstronomy} and {\sc Glee}. The Keck NIRC2 pixel scale is $\sim$50\% smaller and Keck achieves an AO-corrected FWHM that is $\sim$3.4$\times$ smaller than LBT LUCI. However, the LBT LUCI data is deeper than the data presented in our analysis and covers both the $J$- and $K$-bands. For the comparison that follows, we shift the reference point used in our analysis to G1 to be consistent with \citetalias{ecker2026}. Although image A has higher astrometric precision, the light centroid of G1 provides a more physically motivated reference, as the lens model parameters, in particular the offsets of the mass centroids, can be straightforwardly defined relative to G1's light center.



\subsection{Astrometric Discrepancies}
\label{subsubsec:offsets}
The comparison between the model parameters reported in our analysis and in \citetalias{ecker2026} revealed an offset in astrometric positions of the lensed images and lens galaxies (G1 and G2). We use the primary lens galaxy, G1, as a common reference point between the two datasets. The raw root-mean-squared (RMS) of the difference between datasets is measured to be $\sim$33 mas. After applying a simple transformation matrix (including translation and rotation) to the positions from \citetalias{ecker2026} we reduce the RMS to $\sim$14 mas. In this correction, rotation is the dominant factor, measuring $\sim$0.62 degrees counterclockwise when transforming the LBT positions to match Keck. We also apply a scale transformation in addition to translation and rotation to account for potential detector effects. With a scale magnitude of 0.99430 ($\sim$0.6\% size reduction from LBT to Keck), the RMS is reduced to $\sim$9 mas. This residual exceeds the statistical positional uncertainties reported in our analysis and in \citetalias{ecker2026} for all images except image E.
The centroids of the galaxy light profiles are softer than those of point-sources and image E is a relatively low signal-to-noise detection in both datasets. Thus, we also experiment with the removal of G1, G2, and image E positions, calculating the transformation matrix with just the multiple images A--D.  With a rotation of $\sim$0.63 deg. and a scale magnitude of 0.99384, the RMS is reduced to $\sim$5 mas. 
The explainable difference is dominated by extrinsic properties related to the observations and/or data reduction and processing. However, there are clearly intrinsic effects that are not accounted for in our simple transformation matrix. It is likely that there exist some higher-order distortions. 

Despite our best attempts to correct for and reduce the differences between the datasets, there remains at least a 5 mas RMS between the Keck and LBT position measurements. These corrections are entirely arbitrary and do not describe the physical causes for the discrepancy. We maintain confidence in our astrometric solutions due to the quality of the Keck NIRC2 calibrations. The distortion correction presented in \citet[][]{lu2016} has a total accuracy of $\sim$1.1 mas and does an excellent job correcting for any nonlinearities in the detector space.
We do not apply any of the corrections via the transformation matrix mentioned above to the data or parameters from our analysis or from \citetalias{ecker2026}.
Ultimately, this offset in positions does not in turn have a substantial effect on the intrinsic model output parameters, especially when considering the reported parameter statistical uncertainties. 

\subsection{Model Parameters}
\label{subsubsec:ecker-params}
In this comparison, it is important to distinguish between extrinsic parameters (positional coordinates, position angles), which are directly affected by coordinate transformations, and intrinsic parameters (Einstein radius, axis ratio, shear magnitude), which characterize the mass distribution and are invariant to rotation and translation.
Both \citetalias{ecker2026} and this work use the circular definition of $\theta_{\mathrm{E}}$ and parameterize ellipticity and external shear as a magnitude and position angle. The ellipticity position angle is measured E of N, whereas the external shear position angle is counterclockwise from West and thus we subtract 90 from the shear position angle reported in \citetalias{ecker2026} for consistency. 
We apply the same pair-wise comparison done in Section~\ref{sec:results} (defined in Equation~\ref{eq:consistency}) between the model parameters from each of the two models (\texttt{lenstronomy} and {\sc Glee}) for the $K$-band data from \citetalias{ecker2026} and each of the two models from (\threecodes) reported here. This is a total of 40 parameter comparisons, all of which are $<2\sigma$. We show each of the comparisons and the significance of their discrepancy in Figure~\ref{fig:E26_comp}. 

\vspace{0.25cm}
\begin{minipage}{0.99\textwidth}
\makebox[\textwidth][c]{
  \includegraphics[keepaspectratio=true,width=0.75\textwidth]{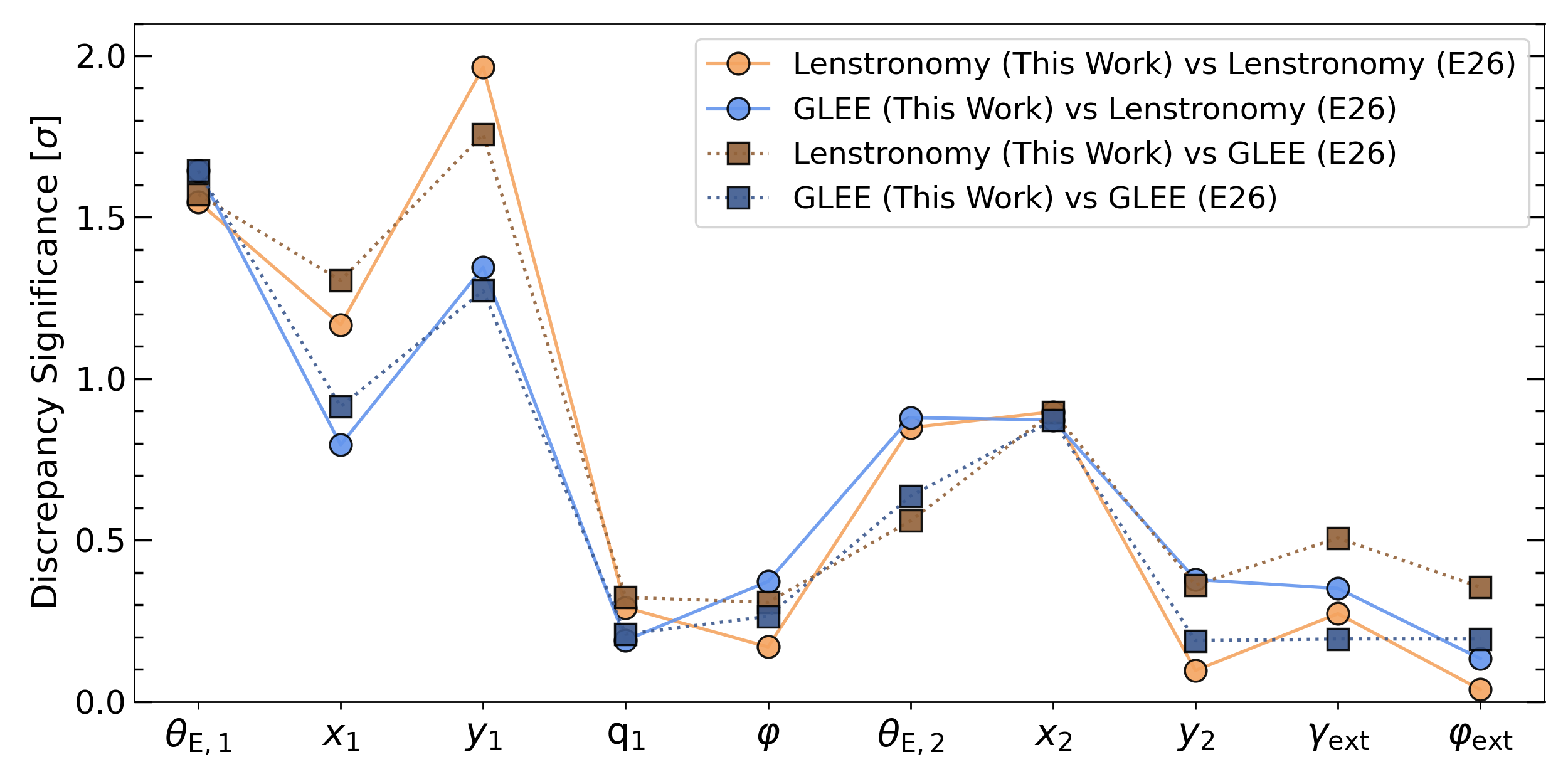}}
\captionof{figure}{The significance of the discrepancy in units of $\sigma$ between the ten lens mass parameters from each of the two lens models reported here and in \citetalias{ecker2026}. This pair-wise comparison metric is defined in Equation~\ref{eq:consistency}. All parameters are consistent to less than 2$\sigma$.}
\label{fig:E26_comp}
\end{minipage}
\vspace{0.25cm}

The parameter with the largest discrepancy is $y_1$, the $y$ position of the primary lens, G1. In particular, the \texttt{lenstronomy} model reported here is in tension with the \threecodes~models reported in \citetalias{ecker2026}. Part of the reason for this is the very tight constraint on $y_1$ from the \texttt{lenstronomy} model in our analysis (see Figure~\ref{fig:posterior}). This very narrow posterior distribution is a consequence of the physical constraint on image multiplicity and not a symptom of the informative priors.
The Einstein radii are intrinsic parameters and are directly affected by any scale change but not by any translational or rotational changes. We see that $\theta_{\mathrm{E,1}}$ is discrepant between both codes in this work and \citetalias{ecker2026} by $\sim1.6\sigma$. Considering the 0.6\% scale difference mentioned above, the Einstein radii are expected to be larger in \citetalias{ecker2026} by $\sim$0\farcs01, which is very close to the 1$\sigma$ uncertainties quoted by both models in both works. Thus, the $\sim1.6\sigma$ discrepancy is explainable in the systematic floor set by the magnitude of the scale transformation. Otherwise, the four comparisons shown in Figure~\ref{fig:E26_comp} show essentially the same pattern pointing towards discrepancies being driven by the systematics of the astrometric inputs and not by differences between the modeling codes.

\section{Conclusions}
\label{sec:conclusions}

In this work, we report the best-fit mass models for the first galaxy-scale strongly lensed supernova with a radial size~$>1\arcsec$, SN~Winny. We leverage exquisite AO imaging data acquired by the NIRC2 instrument on Keck II (K$_p$-band). The AO-corrected seeing for the observations is estimated to be 0\farcs065 on a detector with a pixel scale of 0\farcs009952. This image quality and resolution is unparalleled from the ground and allows for the measurement of precise astrometric positions of the lensed SN images. 
We employ two independent lens modeling codes, \threecodes,~to measure the physical parameters of the lens mass components, which are constrained solely by the observed positions of the five point-like multiple images of SN~Winny. 

This system is appropriately parameterized with a model comprised of a SIE, SIS, and external shear. With twelve free parameters (including two nuisance parameters for the unlensed source position) and ten observables ($2\times5$ image positions) we rely on narrow, well-informed physical priors on the positions of the primary and secondary masses (G1 and G2) based on the centroids of their respective light models. Two independent lens modeling codes, \threecodes, are used to model the total mass distribution of the system, with matched prior ranges and distributions, differing only where the internal parameterization of each code requires it. Beyond minimal differences in the definitions of select parameters (all of which were corrected following modeling), both codes use very similar optimization and sampling techniques. 

Both models are in very good agreement. In our comparison of both models, all ten parameters agree to $<0.6\sigma$. Similarly, predicted magnifications are all consistent to $<0.7\sigma$. The values of the model parameters reported by this work are also consistent to $<2\sigma$ with those reported by \citet[][]{ecker2026}. We do, however, identify small discrepancies in the observed lens galaxy and multiple image positions between the two analyses (Keck and LBT). The optimal transformation includes a translation, rotation, and scale component, with the latter two dominating the correction. Its effect on the intrinsic lens mass model parameters is either not a major concern or explainable when considering the systematic floor set by the magnitude of the scale factor. The scale correction for positions shows a decrease in size from LBT to Keck by $\sim0.6\%$. This difference in scale for $\theta_{\mathrm{E,1}}$ (the most discrepant intrinsic parameter) is comparable to the $1\sigma$ statistical uncertainties reported in this work and \citetalias{ecker2026}.

The lens models reported here and in \citet[][]{ecker2026} are intrinsically limited by the data, preventing more in-depth analyses, including predictions of time delays in pursuit of measuring the value of $H_0$. On the lens modeling front, additional deep, high-resolution imaging like that from the James Webb Space Telescope will be very helpful in constraining the radial shape of the lens mass distributions. Use of more descriptive mass models, such as an elliptical power-law or a composite Navarro–Frenk–White $+$ stellar mass, can be utilized when modeling both SN positions $+$ host galaxy light, jointly \citep{birrer2020a}. Additionally, stellar kinematic analysis can provide independent measurements of the radial total mass profile and, when combined into lens models, has been shown to be additionally constraining on the mass profile \citep{shajib2023a}. In order to measure the time-delay observationally, additional time-series photometry and spectroscopy is needed. As mentioned in \citetalias{ecker2026}, there are a number of parallel efforts using a variety of observatories in order to obtain this data.

SN~Winny is the first galaxy-scale lensed supernova with expected time-delays that are suitable for cosmology. This presents an incredible opportunity to make a precise measurement of $H_0$, a critically complementary measurement to those made using SNe lensed by galaxy clusters \citep[e.g.][]{grillo2024a,pascale2025a}. As we enter the era of the Vera~C.~Rubin Observatory and soon the Roman Space Telescope, the number of ``cosmology-grade", galaxy-scale strongly lensed supernovae will grow rapidly \citep{arendse2024a,murieta2024a}. 
The discovery of SN Winny in a system previously identified as a strong lens candidate \citep{canameras2020a} underscores the value of systematic lens searches \citep[e.g.,][]{jacobs2019b,huang2020a,huang2021a,storfer2024a,storfer2026a} in enabling the rapid identification and follow-up of lensed transients.
This effort will lead to unprecedented progress in constraining H$_0$ with strong lensing, pushing toward a 1\% measurement \citep{qi2022a,birrer2024r}, underscoring the importance of a system like SN~Winny. The strategies developed to characterize galaxy-scale lensed SNe like SN~Winny are being deployed for the first time. As new lensed SNe are discovered, this process will continue to be refined. This includes observational follow-up as well as rapid and robust lens modeling methodology \citep{baltasar2026}. SN~Winny marks the beginning of a new era in lensed SN cosmology, which will continue to explode in the coming years with new observatories and refined strategies.
\clearpage
\begin{acknowledgments}
We thank Leon Ecker, Allan Schweinfurth, and Sherry Suyu for useful discussions and feedback.

Some of the data presented herein were obtained at Keck Observatory, which is a private 501(c)3 non-profit organization operated as a scientific partnership among the California Institute of Technology, the University of California, and the National Aeronautics and Space Administration. The Observatory was made possible by the generous financial support of the W. M. Keck Foundation.

The authors wish to recognize and acknowledge the very significant cultural role and reverence that the summit of Maunakea has always had within the Native Hawaiian community. We are most fortunate to have the opportunity to conduct observations from this mountain.

CJS is supported by NASA Euclid General Investigator Program Grant No. 80NSSC25K7735 and NASA Roman ROSES Grant No. 80NSSC24K0353.

KCW is supported by JSPS KAKENHI Grant Numbers JP24K07089, JP24H00221.

AA acknowledges financial support through the Beatriz Galindo programme and the project PID2022-138896NB-C51 (MCIU/AEI/MINECO/FEDER, UE), Ministerio de Ciencia, Investigación y Universidades.

CG acknowledges financial support through grant PRIN-MIUR 2020SKSTHZ.

KSM is supported by the European Union's Horizon 2020 research and innovation programme under European Research Council Grant Agreement No. 101002652 (BayeSN)

DOJ acknowledges support from NSF grants AST-2407632, AST-2429450, and AST-2510993, NASA grants 80NSSC24M0023 and 80NSSC24K0353, and HST/JWST grants HST-GO-17128.028 and JWST-GO-05324.031, awarded by the Space Telescope Science Institute (STScI), which is operated by the Association of Universities for Research in Astronomy, Inc., for NASA, under contract NAS5-26555.  This work is also funded in part by the Gordon and Betty Moore Foundation through Grant GBMF13900 to D.O.J.

\end{acknowledgments}

\bibliographystyle{aasjournal}
\bibliography{dustarchive}

\end{document}